\documentclass[showpacs,aps,pra,twocolumn]{revtex4}

\usepackage{graphicx}
\usepackage{ulem}
\usepackage{color}
\newcommand{\ket}[1]{|#1\rangle}
\newcommand{\bra}[1]{\langle #1|}

\begin{document}

\title{Quantum correlations and mutual synchronization}
\author{Gian Luca Giorgi}
\author{Fernando Galve}
\author{Gonzalo Manzano}
\author{Pere Colet}
\author{Roberta Zambrini}
\affiliation{IFISC,
Instituto de F\'isica Interdisciplinar y Sistemas Complejos (UIB-CSIC), Campus Universitat Illes Balears,
E-07122 Palma de Mallorca, Spain}
\date{\today}

\begin{abstract}

We consider the phenomenon of  mutual synchronization in a fundamental quantum system of two detuned quantum
harmonic oscillators dissipating into the environment.  
We identify the conditions leading to this spontaneous
phenomenon 
showing that the ability of the
system to synchronize is related to the existence of disparate decay rates and is accompanied by robust quantum
discord and mutual information between the oscillators, preventing the leak of information from the system. 

\end{abstract}


\maketitle

\section{Introduction}
Synchronization phenomena have been observed in a broad range of physical, chemical and biological
systems under a variety of circumstances \cite{Pikovsky}. In some instances, synchronization (also
known as entrainment in this context) is induced by the presence of an external forcing or driving
that acts as a pacemaker; in others it appears spontaneously as a consequence of the interaction
between the elements. The latter case is the most relevant from the  point of view of complexity
since it appears as an emergent phenomenon that takes place despite the natural differences
between the elements. Collective synchronization, whose simplest description can be given in terms
of coupled self-sustained oscillators, is found in relaxation oscillator circuits, networks of
neurons, cardiac pacemaker cells or fireflies that flash in unison \cite{Strogatz}. 
A key ingredient for collective synchronization is dissipation, which
is responsible for collapsing any trajectory of the system in phase space in a lower
dimensional manifold.  

Synchronization has also been studied in the quantum world in the case of  entrainment induced by an external
driving \cite{Hanggi}. Difficulties in addressing quantum collective synchronization come from the fact that in
linear oscillators dissipation will lead to the death of the oscillations after a transient while the extension to
the quantum world of nonlinear phase oscillators does not allow for an insightful treatment.   In this
work, however, we take a  step toward the understanding of quantum collective synchronization,
showing that it is possible to fully characterize synchronization during the transient dynamics. We find that
synchronization happens in the presence of a common bath  due to a separation between dissipation rates.  
Different groups have recently approached this subject considering the (classical) synchronization of
nanoscopic and microscopic systems susceptible of having quantum behavior \cite{Lipson,Marquardt,Milburn}.
In this work we establish the connections between  the phenomenon of
synchronization and quantum correlations  in the system. 

\section{System}

We consider two coupled quantum harmonic oscillators dissipating into the environment 
\cite{PazRoncagliaPRLPRA,Liugoan,hupazang}  with different frequencies \cite{Galve_PRA2010}, which is arguably one
of the most fundamental prototypical models. Current experimental realizations in the quantum regime include
nanoelectromechanical structures and optomechanical devices \cite{NEMS_optomech} as well as separately 
trapped ions whose direct coupling has been reported recently \cite{Wineland-Blatt}. The system Hamiltonian for
$\hbar=1$ and unit masses is
\begin{eqnarray} \label{eq:1}
 H_{S}=\frac{p_1^2}{2}+\frac{p_2^2}{2}+\frac{1}{2}(\omega_1^2 x_1^2+\omega_2^2 x_2^2)+\lambda
x_1 x_2,
\end{eqnarray}
where  $|\lambda|<\omega_1 \omega_2$  (attractive potential) and we allow for frequency diversity.  The free
Hamiltonian is diagonalized by a rotation in the $x_1-x_2$ plane, with  $\theta$ the angle that gives the
eigenvectors $\{{X}_\pm\}$ a function of the coupling  $\tan2\theta={2\lambda}/{\omega_2^2-\omega_1^2}$.   Master
equations for both a common  bath (CB) and separate baths (SB) have been compared by also analyzing entanglement decay
time in Ref. \cite{Galve_PRA2010},  where both the similarity of the frequencies of the oscillators and the coupling
strength were shown to contribute to the preservation of entanglement for a CB, leading to asymptotic entanglement in the case
of identical frequencies \cite{PazRoncagliaPRLPRA,Prauzner,Liugoan}. The transition from SB to one CB underlies the
capability of entanglement generation discussed in Ref. \cite{SB_CB} and a physical implementation of the latter
has been proposed  recently \cite{Paz_CB}.

  Following Ref. \cite{Galve_PRA2010}, the system dynamics is described by a master equation that is valid in
the weak-coupling limit between the system and environment, without the rotating-wave approximation \cite{breuer}. Even if
the obtained master equation has the same form as the exact one  \cite{hupazang}, the coefficients are approximated
for weak coupling $\gamma$ between the system and environment. Taking this equation for strong coupling can lead to
unphysical values for the reduced density  and violation of positivity can appear at low temperatures and for
certain initial states \cite{breuer}. In the following we restrict our analysis  to small $\gamma=0.01\omega_1$
where we never encounter any unphysical dynamics. This is consistent  with the fact that deviations of
this  master equation from one in the Lindblad form (preserving positivity) are in fact small for high temperatures (here
$T=10\omega_1$ in natural units). Particularly useful for the purpose of understanding the physical behavior of the
oscillators dissipation is the master equation in the basis of the normal modes of the system Hamiltonian as given
in Appendix \ref{appA}. 

An important observation is that  our results do not in fact depend on the specific choice of this master equation.
In particular, in  Appendix \ref{appB} we compare our results with the one obtained from a master equation in the Linblad
form, obtained by a rotating-wave approximation. Within this approximation the master equation is 
known to be in
the Lindblad form \cite{breuer,rivas} and we find almost exactly the same results as that obtained with the master
equation (\ref{bas}). Therefore the phenomena predicted in the following do not depend on the  specific details of the
considered master equation.

\section{Synchronization}

The dynamical behavior of the two oscillators can be analyzed through their average
positions,  variances, and correlations, as we deal here with Gaussian states. The presence of a CB or of two
(even if identical) SB leads to different friction terms in the
dynamical equations of both first- and second-order moments  \cite{Galve_PRA2010} with
profound consequences. We recall that for a CB only the sum of  positions $x_+=x_1+x_2$
is actually dissipating and this does not coincide with $X_+$ unless the oscillators are identical.

\begin{figure}[h]
\begin{center}
\includegraphics[width=8cm]{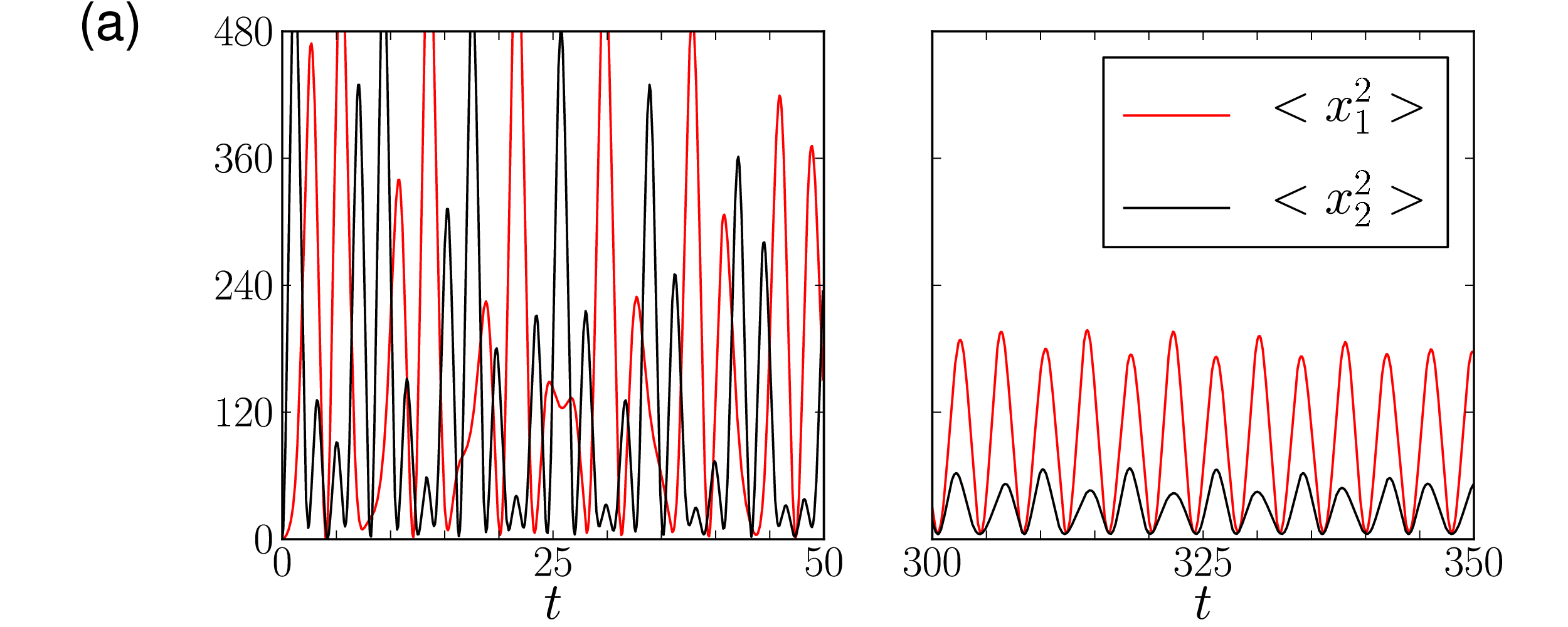}
\includegraphics[width=8cm]{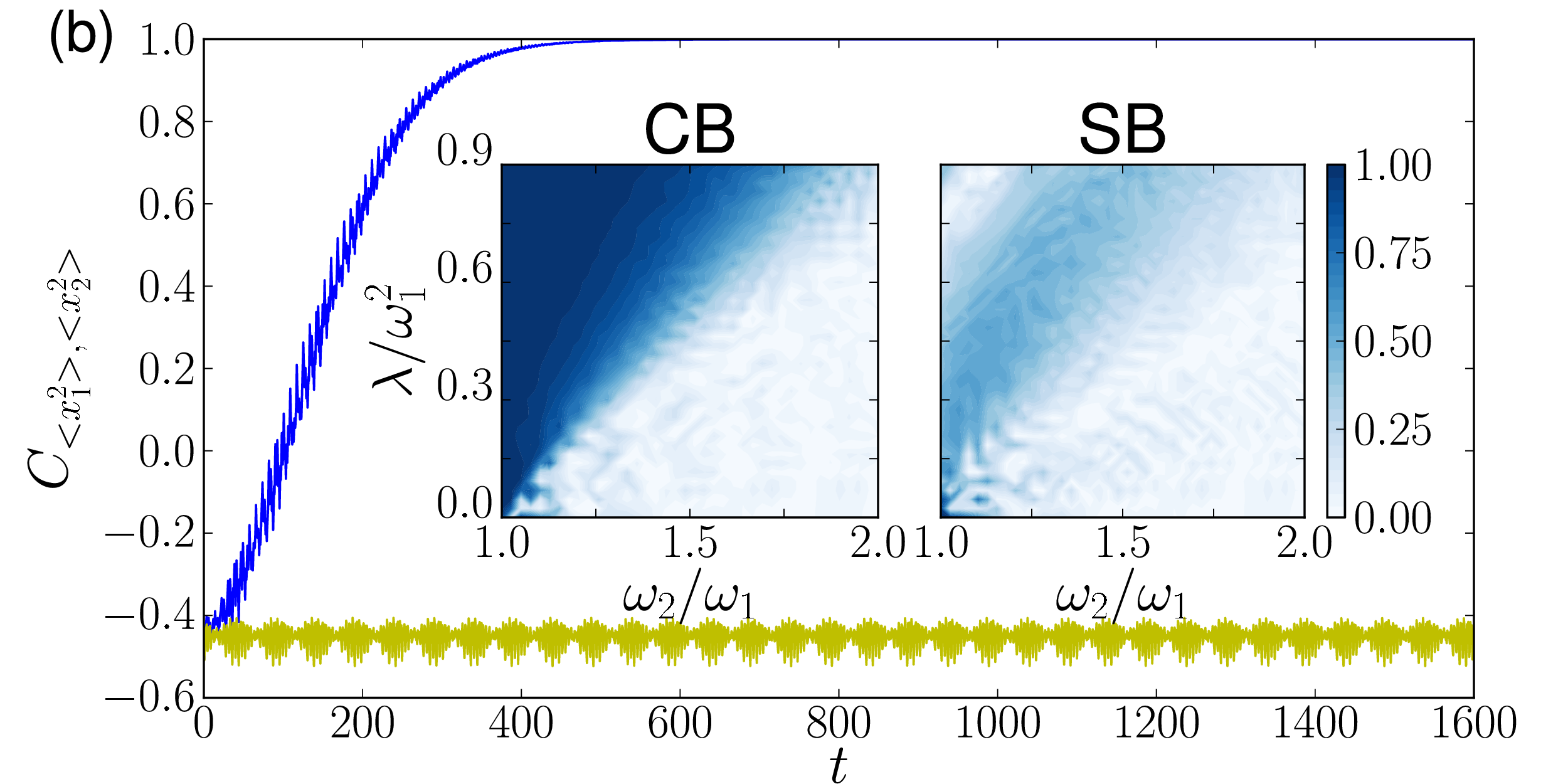}
\end{center}
\caption{ (a) $\langle x^2_{1}(t)\rangle/{\cal S} $ (light color)  and  $\langle x^2_{2}(t)\rangle /{\cal S} $ 
(black) normalized with the shot noise ${\cal S}$, for 
$\omega_2=1.4\omega_1$ 
and $\lambda=0.7\omega_1^2$ starting from squeezed states for  a CB and (b)  synchronization 
$C_{\langle x^2_{1}\rangle\langle x^2_2\rangle}(t,15)$  for a CB  (dark color) and
SB (light color) for temperature $T=10\omega_1$ (in natural units). 
The insets show synchronization values $|C_{\langle x^2_{1}\rangle\langle x^2_2\rangle}|$
varying  $\omega_2/\omega_1$ and 
$\lambda/\omega_1^2$ at $t=300$. Time $t$ is scaled with $\omega_1$ (and is therefore adimensional) and $\gamma=0.01\omega_1$, here and in the following.
The initial state is separable with squeezing parameters 
$2$ and $ 4$, respectively, in the two
oscillators. 
\label{fig1}}
\end{figure}

Figure~\ref{fig1}(a) shows the variance dynamics of two oscillators starting from two vacuum squeezed 
states.  
To quantify synchronization  between two functions $f(t)$ and $g(t)$, we adopted a commonly used
indicator, namely,    $C_{f,g}(t,\Delta t)=\overline{ \delta f \delta g}/\sqrt{\overline{  \delta f^2}
~\overline{\delta g^2} }$ where the bar stands for a time average $\overline{f}=\int_{t}^{t+\Delta
t}dt'f(t')$ with time window $\Delta t$ and $\delta f=f-\overline{ f}$.  For similar evolutions 
$|C|\sim 1$, while $|C|\sim 0$ for different dynamics. The position variances for CB
[Fig.~\ref{fig1}(a)] show a transient dynamics  without any similarity between them, also in antiphase
($C_{\langle x^2_{1}\rangle\langle x^2_2\rangle}< 0$), before reaching full synchronization 
[Fig.~\ref{fig1}(b)].

 A comprehensive analysis shows that  this behavior is robust considering (i) different  initial
conditions, (ii) any second-order moments of the two oscillators (either of positions  $x_{1,2}$  or
momenta $p_{1,2}$ or any arbitrary quadrature), and  (iii) a whole range of couplings and detunings.  Regarding (i), an important
observation is that  while in an isolated system  the dynamics is strongly determined by the initial
conditions, this is not the case in the presence of an environment. After a transient (in which the
initial conditions have an important role), we actually find synchronization  independently of the
initial state, with detuning and  oscillator  coupling therefore being the only relevant parameters. 
The full analysis (iii) for a CB allows us to conclude that synchronization arises faster for nearly
resonant oscillators and that the deteriorating effect of detuning can be \textit{proportionally} 
compensated  for by strong
coupling, as represented in the CB inset of Fig.~\ref{fig1}(b).

Moving now to the case of separate baths, a completely different scenario appears. The quality of the
synchronization is generally poor (small $|C|$), not improving in time and dependent on the initial
condition. The full parameters map for $|C|$ is shown in the SB inset of  Fig.~\ref{fig1}(b).  In
this case the oscillators do not synchronize in spite of their coupling even considering long times
when finally the system thermalizes.

\begin{figure}[h]
\includegraphics[width=4cm]{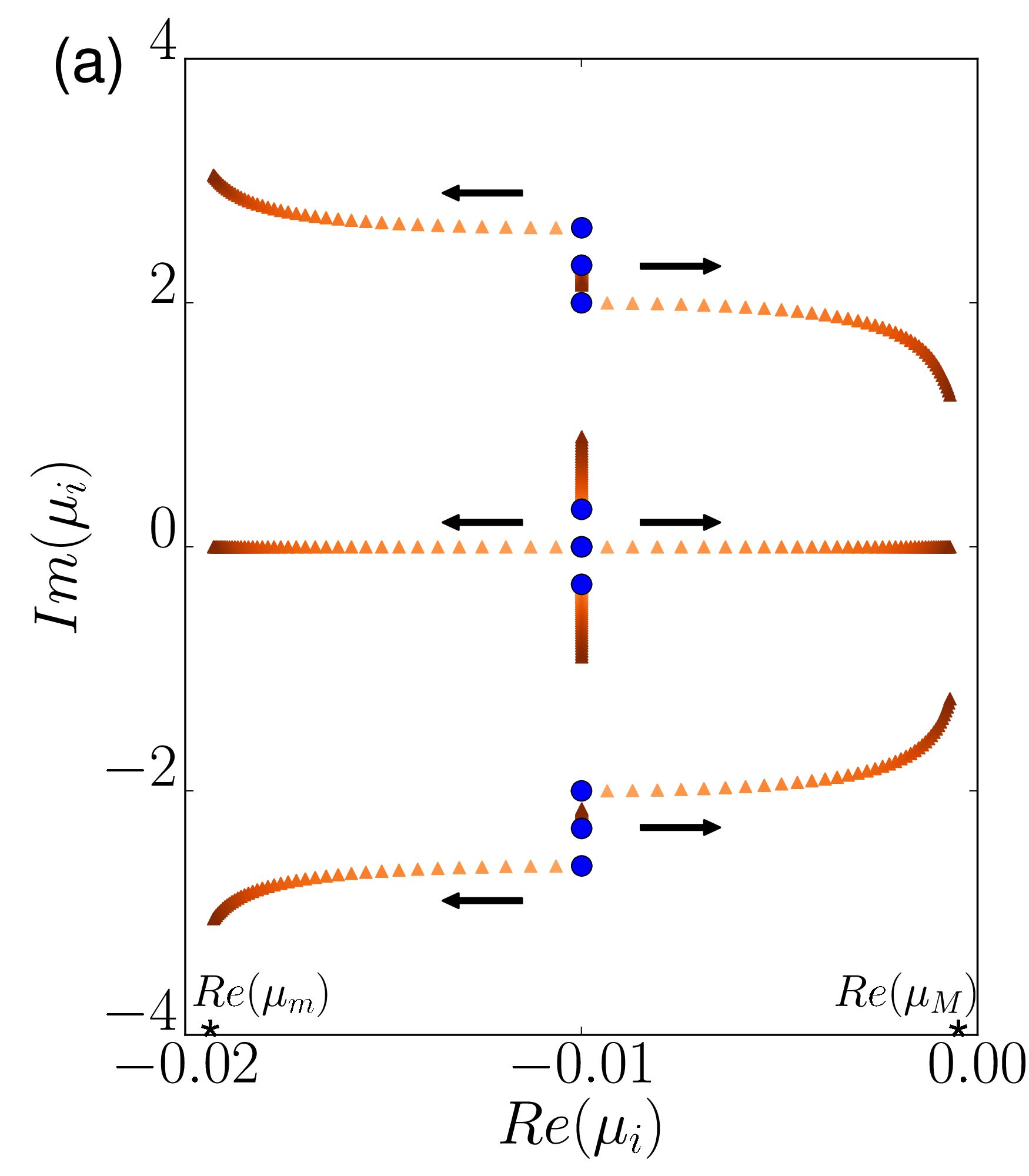}\includegraphics[width=4cm]{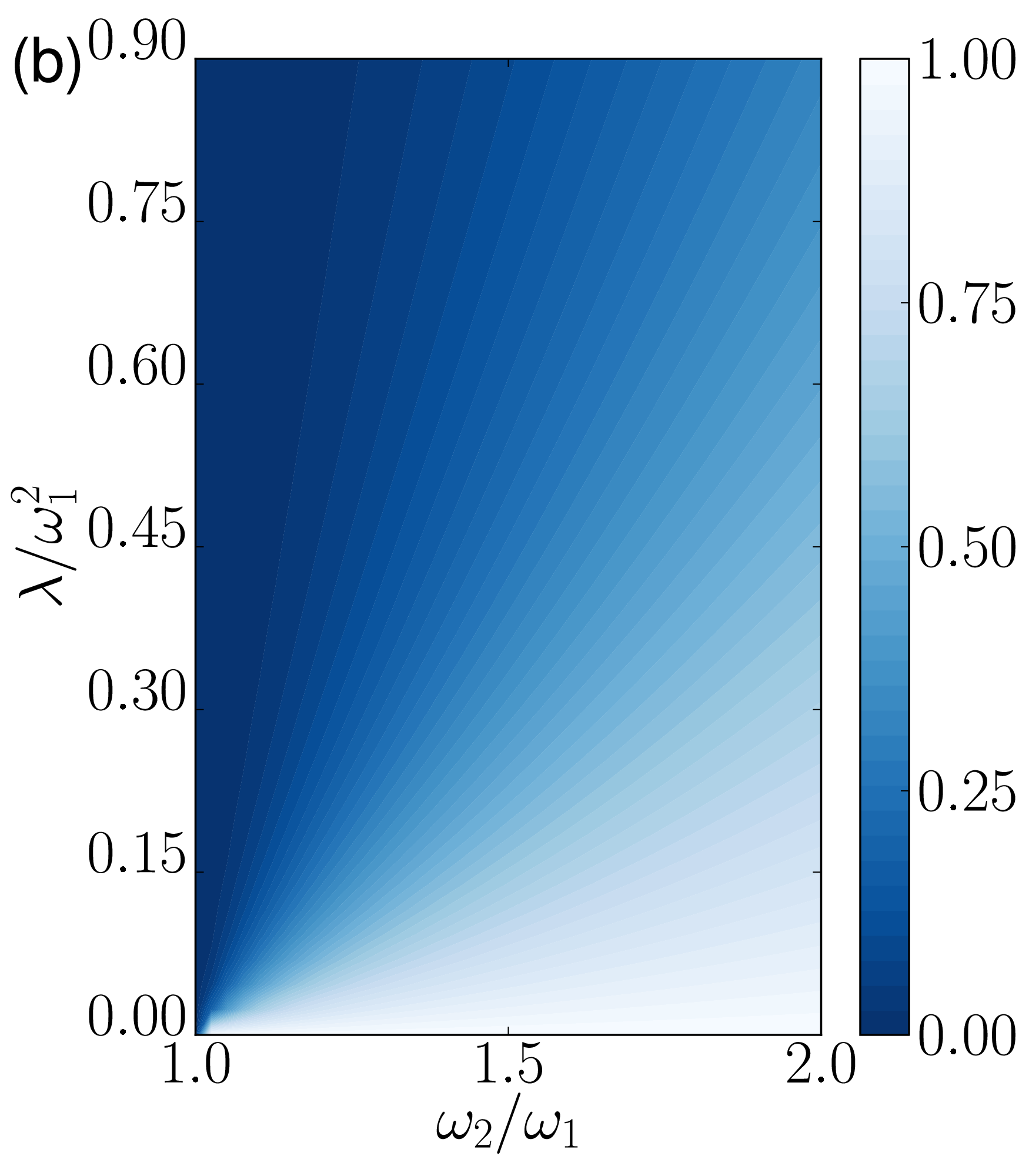}
\caption{ (a) Eigenvalues $\mu_i$ in the complex plane for a CB for $\omega_2/\omega_1 =1.31$ and
increasing the coupling from  $\lambda=0$ [circles with ${\rm Re}(\mu_i)\sim-0.01$] to
$\lambda=0.9 \omega_1^2$ in the direction of darker colors.  (b)  Ratio between minimum and maximum
eigenvalue ${\rm Re}(\mu_{m})/{\rm Re}(\mu_{M})$ for a CB. \label{fig2}}
\end{figure}

 The appearance of a synchronous dynamics only for a CB can be understood by considering the  time evolution
of the second-order moments. The matrix  governing their time evolution  \cite{Galve_PRA2010} (see also 
 Appendix~\ref{appA})  has 
complex eigenvalues $\{\mu_i\}$ ($i=1,..,10$), named in the following \textit{dynamical eigenvalues}.
Their real and imaginary parts determine the  decays and oscillatory dynamics of all second-order
moments and  variances. As shown in Fig.~\ref{fig2}(a), when $\lambda=0$ all the eigenvalues are along
the line $-0.01$ and for increasing  coupling in the case of one CB they move in the
complex plane assuming three different real values.
In contrast, for SB all dynamical eigenvalues have similar real parts that remain almost unchanged
when varying parameters. Hence for   SB the ratio between the maximum and the minimum eigenvalues  ${\rm
Re}(\mu_{M})/{\rm Re}(\mu_{m})\approx 1$ is almost constant for all parameters,  while for a CB $and$ for
parameters for which synchronization is found 
[CB inset in Fig.~\ref{fig1}(b)] $ {\rm Re}(\mu_{M})/{\rm Re}(\mu_{m})<<1 $,  as shown in
Fig.~\ref{fig2}(b). 
In this parameters regime, after a transient time, only the least-damped eigenvector survives, thus  fixing the
frequency of the  whole dynamics of the  moments.  As a consequence of this mechanism, synchronization
is observed by considering the expectation values of any quadrature of the oscillators as well as higher-order
moments.

We obtain an approximated analytical estimation of time scales by considering the master equation in the eigenbasis
$\{{X}_\pm,{P}_\pm\}$ of the free Hamiltonian [Eq.~(\ref{eq:1}],   as detailed in Appendix~\ref{appA}.
The master equations for common and separate baths have the same expression  in the case of detuned oscillators
and the nature of dissipation (a CB or SB)  appears only   in the form of the damping coefficients. 
By eliminating the oscillating terms in the dynamics in the interaction picture, one obtains
that, within this approximation, the decay rates of $\left<{P}_\pm^2\right>$  are given by
 \begin{eqnarray} \label{eq:decaysSB}
 \tilde   \Gamma_{\pm\pm}^{SB}&=&c^2 \Gamma_{11}+s^2  \Gamma_{22}\pm cs \Gamma_{12} 
\end{eqnarray} 
for SB and  
\begin{eqnarray} 
\label{eq:decaysCB}  
\tilde\Gamma_{\pm\pm}^{CB}=(c\pm s)(c\Gamma_{11}\pm s\Gamma_{22})+(1\pm2sc)\Gamma_{12} 
\end{eqnarray} 
for a CB, where $c=\cos\theta$, $s=\sin\theta$ (with $\theta$  previously defined diagonalization angle of
$H_S$), and $\Gamma_{11,22,12}$  appear  in the original master equation (see the Appendix of Ref.
\cite{Galve_PRA2010}). 
These approximated decays for the variances, together with  their  average $(\tilde  
\Gamma_{--}+\tilde   \Gamma_{++})/2 $ (for $\left<{P}_+{P}_-\right>$), for a CB and SB do agree very 
well with the real parts of the dynamical eigenvalues.

As mentioned before, synchronization (for a CB) is observed by looking at  the dynamics of  both first- and second-order
moments: the ratio between minimum and maximum  dynamical eigenvalues  is the same in both
cases. Still, our interest is in the second-order moments due to their relevance in the quantum information shared
by the oscillators. Furthermore, inspection of first-order moment dynamics allows us to establish connections
with what is known in classical systems \cite{Pikovsky}. Two studied scenarios for classical synchronization are
the diffusive coupling where both oscillator dampings depend on the difference of the velocity  and the 
direct coupling where each one depends on the velocity of the other \cite{Pikovsky}. The quantum harmonic
oscillators  considered here for a CB display in their first-order moments a diffusive coupling up to a change of
sign, which  explains the {\it antiphase} character of their synchronization.

\begin{figure}[h]
\includegraphics[width=9cm]{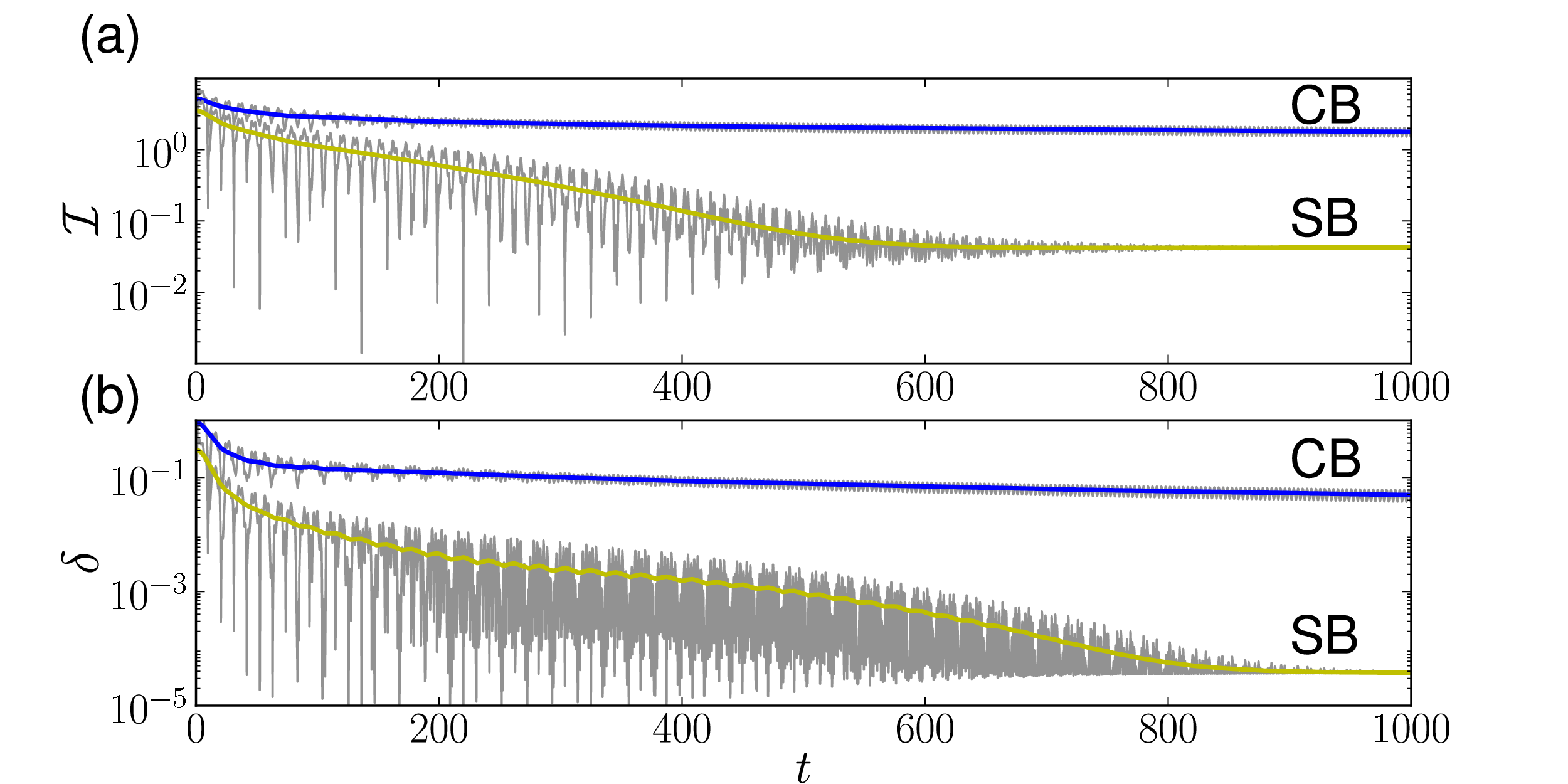}
\caption{   (a) Mutual information and (b) discord  on a logarithmic scale for a common bath and separate
baths  (SB). Both the exact time evolutions and the filtered ones (Gaussian filter) are shown.
The parameters are $\omega_2=1.05\omega_1$  and $\lambda=0.3\omega_1^2$.
\label{fig4}}
\end{figure}

\section{Quantum correlations}

Once the conditions for synchronization  to arise have been established, we explore this phenomenon by focusing on
the information aspects, through  mutual information shared by the oscillators and their quantum
correlations. In particular, the total correlations between the oscillators are measured by the
mutual information ${\cal{I}}(\varrho)=S(\varrho_1)+S(\varrho_2)-S(\varrho)$ where $S$ stands for the
von Neumann entropy, and $\varrho_{1(2)}$ is the reduced density matrix of each harmonic oscillator. 
A possible partition of correlations into quantum and classical parts that has received much
attention  recently is given by  the quantum discord \cite{zurek,vedral}
$\delta(\varrho)=\min_{\{\Pi_i\}}\left[S(\varrho_2)-S(\varrho)+S(\varrho_1|\{\Pi_i\})\right]$,  with
the conditional entropy defined as $S(\varrho_1|\{\Pi_j\})=\sum_ip_iS(\varrho_{ 1|\Pi_i})$,
$\varrho_{1|\Pi_i}= \Pi_i \varrho \Pi_i /{p_i} $ density matrix after a complete
 measurement $\{\Pi_j\}$ on the second oscillator, and  $p_i={\rm
Tr}_{12}(\Pi_i\varrho)$.  The importance of
this measure of the quantumness of
correlations relies on its capability  to distinguish and  understand classical and quantum
behaviors \cite{discord}. Quantum discord has been reported recently also for continuous variables
in Gaussian states \cite{paris-adesso}.

Dissipation degrades all  quantum and classical correlations \cite{ijqi}.  Nevertheless,
important differences are found when comparing a CB and SB for the same
parameter choices. In Fig.~\ref{fig4} we show a fast decay of all  the total [Fig.~\ref{fig4}(a)] and
 quantum [Fig.~\ref{fig4}(b)] correlations for SB. In contrast, for a CB we find that
after a short transient  both mutual information and discord 
oscillate around an  almost constant value and their decay is nearly frozen.  
 For these parameters and a common environment the oscillators
synchronize and $C_{\langle x^2_{1}\rangle\langle x^2_2\rangle}=0.95$ at $t\sim 270$.  
The robustness of the quantum correlations for long times 
 in synchronizing oscillators in a CB and the deep differences
with the case of SB is surprising also because their respective asymptotic values
are really similar for detuned oscillators. In other words, the upper CB curve in Fig.~\ref{fig4}(a) [or Fig.~\ref{fig4}(b)] will 
eventually thermalize, converging to a value very similar the one obtained for SB, while strong differences in the
asymptotic values actually appear only in the case of identical oscillators \cite{PazRoncagliaPRLPRA}. 
As a further result, the effect of increasing the temperature is mostly on the asymptotic state while
the main features of the   dynamics  described here are still observed.
 
\begin{figure}[h]
\begin{center}
\includegraphics[width=8cm]{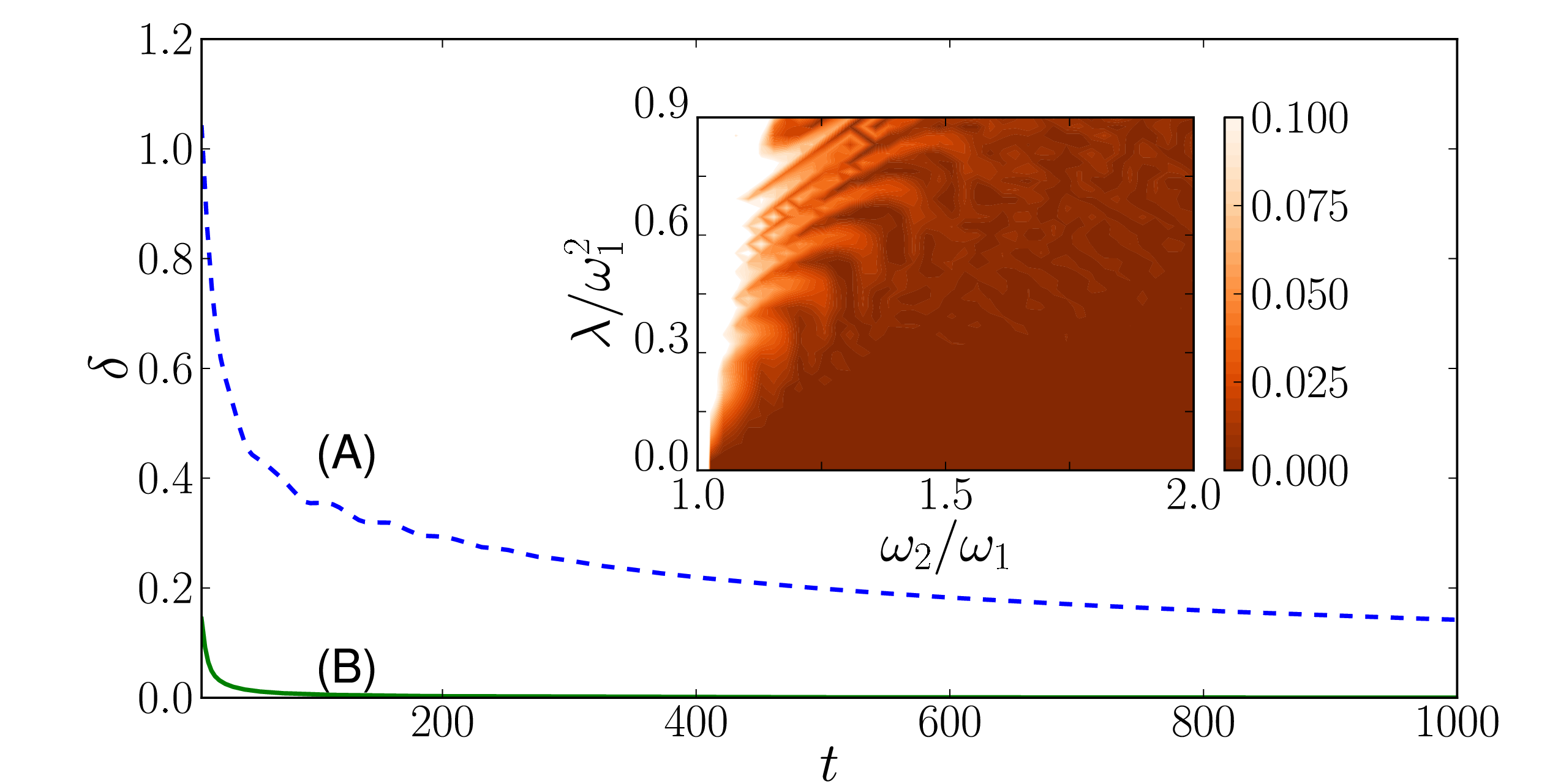}
\end{center}
\caption{ Evolution of the discord for a CB and  the parameters   $\omega_2/\omega_1= 1.05$ and
$\lambda=0.8\omega_1^2$ (dashed line A)  and   $\omega_2/\omega_1= 1.4 $ and $\lambda=0.3\omega_1^2$ 
(solid line B). The inset represents the quantum discord at  $t=300$ for a CB.\label{fig5}}
\end{figure}

We now focus on the case of a CB to look for specific quantum features of the synchronization
exploring different parameters regimes.  The comparison of mutual information and discord in cases in
which there is synchronization or  the system dissipates without having time to synchronize is given
in Fig.~\ref{fig5} (upper and lower curves, respectively) where we filter out the fast oscillations
to highlight the decay dynamics. For small  coupling  and large detuning, discord (shown in
Fig.~\ref{fig5} for $\lambda/\omega_1^2=0.3$ and $\omega_2/\omega_1=1.4$) and mutual information are
rapidly degraded. In  this case, when $t=200$ there is no synchronous dynamics and $C_{\langle
x^2_{1}\rangle\langle x^2_2\rangle}\sim 0$. In contrast, for strong coupling or for small
detuning,  synchronization occurs fast: for $\lambda/\omega_1^2=0.8$ and $\omega_2/\omega_1=1.05$,
$C_{\langle x^2_{1}\rangle\langle x^2_2\rangle}(t=200)\sim 1$. In this case, after a short transient,
the  dynamics of discord is  almost frozen and it remains $robust$ against decoherence.  In exploring
different parameter regimes we conclude that fast decay of classical and quantum correlations  is
found in cases in which there is no synchronization while the emergence of synchronization
accompanies robust correlations against dissipation  (frozen decay). The inset in Fig.~\ref{fig5}
represents the value of the discord after the fast decay (here for $t=300$), where it is expected to
be already in the plateau. There is a rather suggestive similarity to the synchronization
map for a CB shown in the inset of Fig.~\ref{fig1}(b).  Considering that also entropy
 shows in this regime a slow growth, we conclude that  synchronized
oscillators are characterized by a reduced leakage of information into the environment.

One might wonder if the presence of a synchronous dynamics has any effect on entanglement, as  in
contrast with pure states mixed states with large quantum correlations can have even vanishing
entanglement \cite{Galve_PRA2011}.  The presence of the environment for
oscillators with different frequencies leads to a complete loss of entanglement in finite short times 
unless the couplings to the CB are "balanced" \cite{Galve_PRA2010}. In the general case of detuned oscillators, even for large
couplings,  entanglement decay is typically faster than the time scales at which  the system reaches
synchronous dynamics both for a CB and SB, mostly at this temperature ($T=10\omega_1$).  Still, longer
survival times for entanglement in a CB are found for small detunings and strong couplings
\cite{Galve_PRA2010}.

\section{Dependence on initial conditions}

We mentioned before that initial conditions do not play any important role in the
appearance of synchronization. Indeed synchronous dynamics of the moments appears when an
eigenmode dominates because of its slow dissipation rate and this goes beyond the specificity of
the choice of the initial state. However, the details of the dynamics do depend on the latter as
we illustrate for the following initial conditions:\\
(i) the separable vacuum state: $\rho=\ket{0}\bra{0}\otimes\ket{0}\bra{0}$ \\
(ii) the two-mode squeezed states
\begin{eqnarray*}
\rho&=&U_{12}(r)\left(\ket{0}\bra{0}\otimes\ket{0}\bra{0}\right)U_{12}^\dagger(r)
\end{eqnarray*}
where $U_{12}(r)=\exp{\left[-r(a_1^\dagger a_2^\dagger-a_1a_2)/2\right]}$ and 
 $a_i (a_i^\dag)$ are the usual annihilation (creation) operators; and
(iii) the separable squeezed state: 
\begin{eqnarray*}
\rho&=&U_1(r_1)\ket{0}\bra{0}U_1^\dagger(r_1)\otimes U_2(r_2)\ket{0}\bra{0}U_2^\dagger(r_2),
\end{eqnarray*}
with $U_i(r_i)=\exp{\left[-r(a_i^{\dagger 2}-a_i^2)/2\right]}$.

\begin{figure}[h]
\includegraphics[width=8cm]{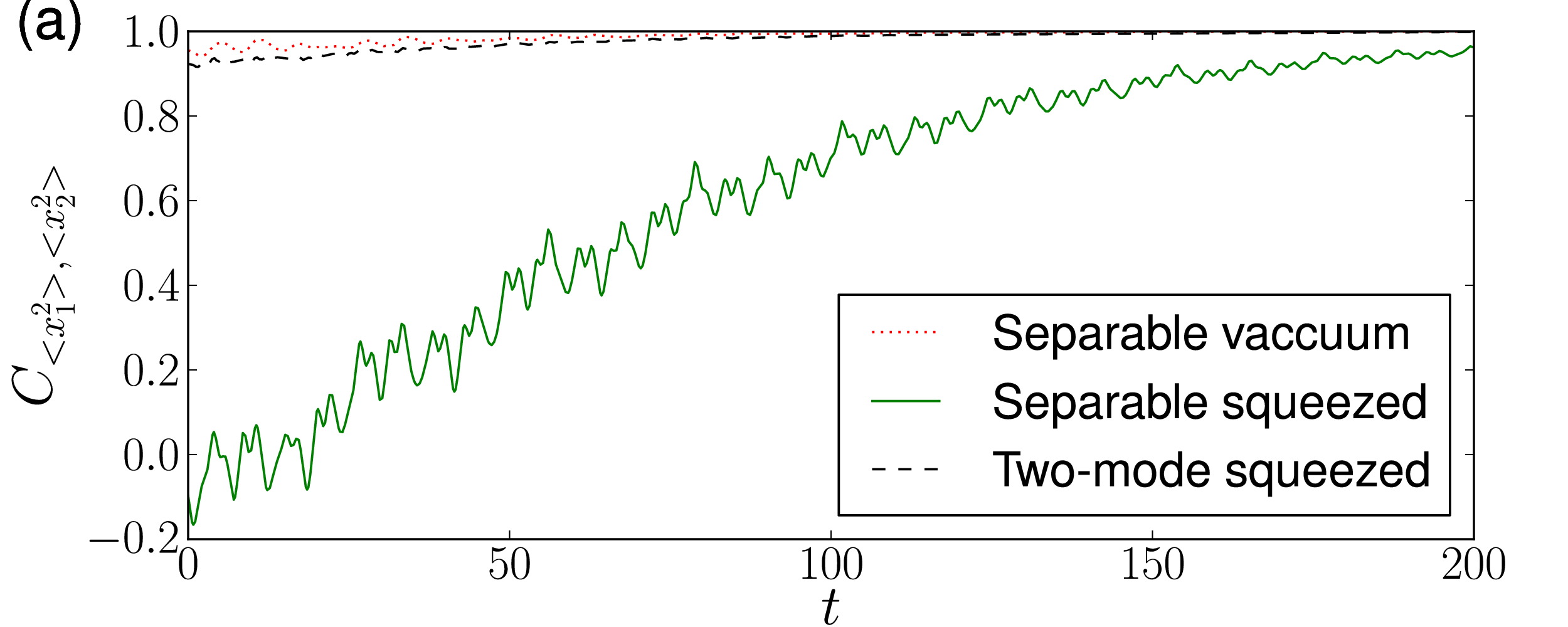}
\includegraphics[width=8cm]{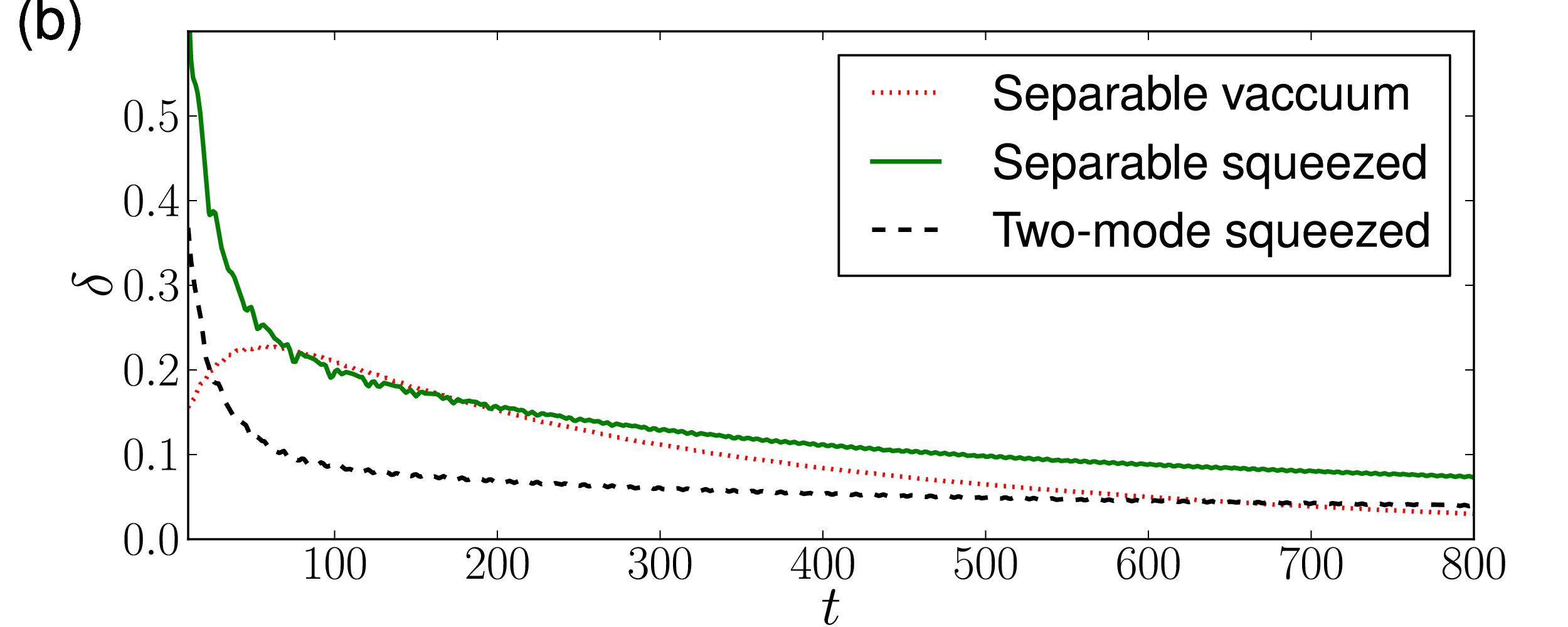}
\caption{ (a) Synchronization indicator and (b) decay of quantum correlations 
 for different initial conditions in the case of a common bath: 
separable squeezed state with squeezings $r_1=2$ and $r_2=4$ (solid line),
 separable vacuum state (dotted line), and  entangled two-mode squeezed state (dashed line) 
 with squeezing $r=2$. Other parameters  are $\omega_2/\omega_1= 1.1$ and 
 $\lambda=0.8\omega_1^2$  }
\label{fig6}
\end{figure}

Instead, quantum correlations $\delta$ depend on the initial condition in 
the sense that more or less of the latter will be present. However after the 
short transient they always reach a plateau where information leakage to the 
environment is greatly reduced. Both information leakage reduction and 
synchronization are part of the same underlying phenomenon: that of a 
dissipation channel being much slower than the other. This behavior is seen 
in Fig. \ref{fig6} where quantum correlations and the synchronization 
indicator are displayed for different initial conditions. We must further 
stress here that, since the asymptotic thermal state has $\delta\sim 10^{-4}$,
 the plateau is expected to be very long.

\section{Conclusions}
Our analysis of the dynamics of dissipative quantum harmonic oscillators has allowed us to  establish under which
conditions synchronization appears. This phenomenon can appear in rather different forms but, in this paper it was reported during the transient dynamics of
a (quantum or classical) system coupled to an environment and relaxing toward equilibrium.
The emergence of synchronization was explained in terms of different
temporal decays governing the system evolution  and related to a separation between the eigenvalues of the
matrix generating the dynamics. We traced synchronization between second-order moment evolution from the
existence of a slowly decaying eigenmode and found approximated expressions for the variance decay
coefficients in very good agreement with the real parts of the  dynamical eigenvalues.  We found that 
synchronization arises  in the presence of a common bath, but not for separate ones, for any strength of the
coupling between oscillators.  It could be of interest to study the transition between such different
scenarios \cite{SB_CB}. 
The relevant parameters are $\lambda/\omega_1^2$ and
$\omega_2/\omega_1$ since the dependence on initial conditions is actually weak.  

We have then characterized mutual synchronization from a quantum information perspective through mutual
information and quantum discord exploring their dynamics in different parameters regimes. We found that there is a
signature of synchronization in the information shared by the oscillators. We have shown that discord and mutual
information are more robust when the oscillators synchronize. In spite of the fact that after thermalizing the
asymptotic discord is negligible for both a CB and SB,  the decay toward this value is clearly frozen in  the presence
of synchronization. In this case total and quantum correlations  display a very slow decay  (plateau) and the
leak of information into the bath is reduced.  The identification of the conditions for the occurrence of
synchronization and its connection with quantum correlations reported here provides the path towards future
extensions such as the study of arrays and networks, the analysis of the role of different environments, or the
exploration of eventual connections with biological systems  in which synchronization is a widespread phenomenon.

\acknowledgments{
We acknowledge financial support from the MICINN (Spain) and FEDER (EU) through
projects FIS2007-60327 (FISICOS) and FIS2011-23526 (TIQS), 
from CSIC through project CoQuSys (200450E566) and from
the Govern Balear through project AAEE0113/09. 
GLG and FG are supported by Juan de la Cierva and JAE programs.}

\appendix

\section{Master equation in the normal modes basis}
\label{appA}

The master equations describing the evolution of the reduced density matrix of the system of
two different oscillators,  up to  second order in the coupling strength, have been
reported in Ref.~\cite{Galve_PRA2010}  for both a common and separate baths. Here we provide
these equations in the basis of the eigenmodes of the Hamiltonian. { We stress that the
exact master equation at all coupling orders has the same structure as ours, the difference
being in the form of its coefficients \cite{breuer,hupazang}. For weak coupling this
equation is a very good approximation to the exact one.  Furthermore, in Appendix
\ref{appB} we show that   the full evolution almost perfectly matches that of a  master
equation obtained by a rotating-wave approximation, the latter being always  positive due to
its Lindblad form.}

\subsection{Common bath}

In the case of a common bath, we assume for the interaction Hamiltonian between the system and the environment
the form $H_I^{CB}=\sum_{k}
g_{k} (b_{k}+b_{k}^\dagger)x_+$, where $x_+=x_1+x_2$, $b_{k}(b_{k}^\dagger)$ creates (annihilates) an
excitation (with energy $\Omega_{k}$) over the $k$th mode of the bath, and where the coupling
coefficients $g_k$ are related to the density of states of the bath $J(\Omega)$ through
$J(\Omega)=\sum_{k}\left(g_{k}^{2}/ \Omega_{k}\right)\delta(\Omega-\Omega_k)$. We assume an Ohmic
environment with a Lorentz-Drude cut-off function, whose spectral density is
\begin{equation}\label{JOhm}
 J(\Omega)=\frac{2 \gamma}{\pi}\Omega\frac{\Lambda^2}{\Lambda^2+\Omega^2}.
\end{equation} 
The  results represented in this work are obtained with
bath-system coupling  $\gamma=0.01 \omega_1$ and cutoff $\Lambda=50\omega_1$.

The eigenvectors of $H_S$ are
\begin{eqnarray}\label{eigenm1}
 X_- &=&\cos\theta x_1-\sin\theta x_2,\\\label{eigenm2}
 X_+ &=& \cos\theta x_2+\sin\theta x_1,
\end{eqnarray} 
with $\tan2\theta={2\lambda}/{\omega_2^2-\omega_1^2}$.
The system eigenfrequencies
$\Omega_{\pm}$ are always different due to the coupling $
{2}\Omega_{\pm}^2=\omega_1^2+\omega_2^2 \pm \sqrt{4\lambda^{2}
+(\omega_2^2-\omega_1^2)^2}$.
Neglecting energy renormalization, the master equation in this eigenvector basis reads
\begin{widetext}
\begin{eqnarray}
\frac{d \rho(t)}{dt}=-i[H_S,{\rho}(t)]-\frac{ \tilde{D}^{CB}_{--}}{2}[X_-,[X_-,\rho]]-
\frac{\tilde{D}^{CB}_{++}}{2}[X_+,[X_+,\rho]]-\tilde{D}^{CB}_{+-} [X_+, [X_-,\rho]] \nonumber\\
+\frac{ \tilde{F}^{CB}_{--}}{2}[X_-,[P_-,\rho]]+\frac{ \tilde{F}^{CB}_{++}}{2}[X_+,[P_+,\rho]]
+\frac{ \tilde{F}^{CB}_{+-}}{2}[X_+,[P_-,\rho]]+\frac{ \tilde{F}^{CB}_{-+}}{2}[X_-,[P_+,\rho]] \nonumber\\
-i\left(\frac{\tilde{\Gamma}^{CB}_{--}}{2}[X_-,\{P_-,\rho\}]+
\frac{\tilde{\Gamma}^{CB}_{++}}{2}[X_+,\{P_+,\rho\}]+\frac{\tilde{\Gamma}^{CB}_{+-}}{2}[X_+,\{P_-,\rho\}]
+\frac{\tilde{\Gamma}^{CB}_{-+}}{2}[X_-,\{P_+,\rho\}]\right),\label{bas}
\end{eqnarray}\end{widetext}
with
\begin{eqnarray} 
 \tilde{D}^{CB}_{--}&=&(c-s)(c D_{11}-sD_{22})+(1-2sc)D_{12}\nonumber\\
\tilde{D}^{CB}_{++}&=&(c+s)(c D_{11}+sD_{22})+(1+2sc)D_{12}\nonumber\\
\tilde{D}^{CB}_{+-}&=&\frac{c^2-s^2}{2}(D_{11}+D_{22}+2D_{12})+sc(D_{11}-D_{22})\nonumber\\
 \tilde{F}^{CB}_{--}&=&(c-s)(c F_{11}-sF_{22})+(1-2sc)F_{12}\nonumber\\
 \tilde{F}^{CB}_{++}&=&(c+s)(c F_{11}+sF_{22})+(1+2sc)F_{12}\nonumber\\
\tilde{F}^{CB}_{-+}&=&(c-s)(c F_{22}+sF_{11})+(c^2-s^2)F_{12}\nonumber\\
\tilde{F}^{CB}_{+-}&=&(c+s)(c F_{11}-sF_{22})+(c^2-s^2)F_{12}\nonumber\\
 \tilde{\Gamma}^{CB}_{--}&=&(c-s)(c \Gamma_{11}-s\Gamma_{22})+(1-2sc)\Gamma_{12}\nonumber\\
 \tilde{\Gamma}^{CB}_{++}&=&(c+s)(c \Gamma_{11}+s\Gamma_{22})+(1+2sc)\Gamma_{12}\nonumber\\
\tilde{\Gamma}^{CB}_{-+}&=&(c-s)(c \Gamma_{22}+s\Gamma_{11})+(c^2-s^2)\Gamma_{12}\nonumber\\
\tilde{\Gamma}^{CB}_{+-}&=&(c+s)(c \Gamma_{11}-s\Gamma_{22})+(c^2-s^2)\Gamma_{12}
\end{eqnarray}
where $c=\cos\theta$, $s=\sin\theta$, and  the dissipation ($\Gamma_{i,j}$) and diffusion ($D_{i,j},F_{i,j}$) coefficients 
are defined in the Appendix of { Ref. \cite{Galve_PRA2010},} specifically in Eqs. (A18)-(A20).
The related set of equations of motion is
 \begin{eqnarray}\label{eqofm}
 \frac{d \langle X_{i}X_{j}\rangle}{dt}&=&\frac{1}{2}\left(\{X_i, P_j\} +\{X_j, P_i\}\right)\\
 \frac{d \langle P_{i}P_{j}\rangle}{dt}&=&-\frac{1}{2}(\Omega_{i}^{2}\langle \{X_{i},P_{j}\} 
 \rangle+\Omega_{j}^{2}\langle \{X_{j},P_{i}\} \rangle)\nonumber\\
&-&( \tilde{\Gamma}^{CB}_{i,i}+ \tilde{\Gamma}^{CB}_{j,j} )\langle P_{i}P_{j}\rangle
- \tilde{\Gamma}^{CB}_{i, -i} \langle P_{j}P_{-i}\rangle \nonumber\\
&-& \tilde{\Gamma}^{CB}_{j,-j} \langle P_{i}P_{-j}\rangle    +\tilde{D}^{CB}_{i,j}\label{eqofm2}\\
\frac{d \langle \{X_{i},P_{j}\} \rangle}{dt}&=&2 \langle P_{i}P_j\rangle-2\Omega_{j}^{2}\langle  X_{i}X_{j}  
\rangle +\tilde{F}^{CB}_{i,j}\nonumber\\ 
&-& \tilde{\Gamma}^{CB}_{j,j}\langle \{X_{i},P_{j}\} \rangle- \tilde{\Gamma}^{CB}_{j,-j} \langle X_{i} P_{-j}  
\rangle \label{eqofm3}
 \end{eqnarray}
where $i,j=+,-$.

The time evolution of the vector ${\bf{R}}$
of all ten moments can be written in a compact matrix form
\begin{eqnarray}
\dot{\bf{R}}= \mathcal{M}{\bf{R}}+{\bf N}.
\end{eqnarray}
The complex eigenvalues of $\mathcal{M}$  are $\{\mu_i\}$ with $i=1...10$,  elsewhere referred as
dynamical eigenvalues.

\subsection{Separate baths}

When the two environments are thought to be identical and independent from each other, the interaction Hamiltonian becomes
\begin{equation}
H_{I}^{SB}=\sum_{i=1}^2\sum_{k}g_{k} (b_{k}^i+b_{k}^{i\dagger}) x_i
\end{equation}
where the annihilation (creation) operators $b_{k}^i(b_{k}^{i\dagger})$ belong, respectively, to the $i$th thermal
bath. The density of states of both of them is that of Eq. (\ref{JOhm}), with the same
 $\gamma$ and  $\Lambda$
introduced before. The master equation has the same structure of Eq. (\ref{bas}) but the
coefficients are modified as follows:
\begin{eqnarray}
 \tilde{D}^{SB}_{--}&=&c^2 D_{11}+s^2 D_{22}-2cs  D_{12}\nonumber\\
\tilde{D}^{SB}_{++}&=&c^2 D_{11}+s^2 D_{22}+2cs  D_{12}\nonumber\\
\tilde{D}^{SB}_{+-}&=&cs(D_{11}-D_{22})+(c^2-s^2)D_{12}\nonumber\\
 \tilde{F}^{SB}_{--}&=&c^2 F_{11}+s^2 F_{22}-2cs  F_{12}\nonumber\\
 \tilde{F}^{SB}_{++}&=&c^2 F_{11}+s^2 F_{22}+2cs  F_{12}\nonumber\\
\tilde{F}^{SB}_{-+}&=&\tilde{F}^{SB}_{+-}=cs(F_{11}-F_{22})+(c^2-s^2)F_{12}\nonumber\\
 \tilde{\Gamma}^{SB}_{--}&=&c^2 \Gamma_{11}+s^2 \Gamma_{22}-2cs  \Gamma_{12}\nonumber\\
 \tilde{\Gamma}^{SB}_{++}&=&c^2 \Gamma_{11}+s^2 \Gamma_{22}+2cs  \Gamma_{12}\nonumber\\
\tilde{\Gamma}^{SB}_{-+}&=&\tilde{\Gamma}^{SB}_{+-}=cs(\Gamma_{11}-\Gamma_{22})+(c^2-s^2)\Gamma_{12}
\end{eqnarray}
Once these coefficients are used instead of those coming from a common bath, the equations of motion 
are formally identical to those of Eqs. (\ref{eqofm})-(\ref{eqofm3}).

\section{Independent decay rates} 
\label{appB}

\begin{figure}[h]
\begin{center}
\includegraphics[width=8cm]{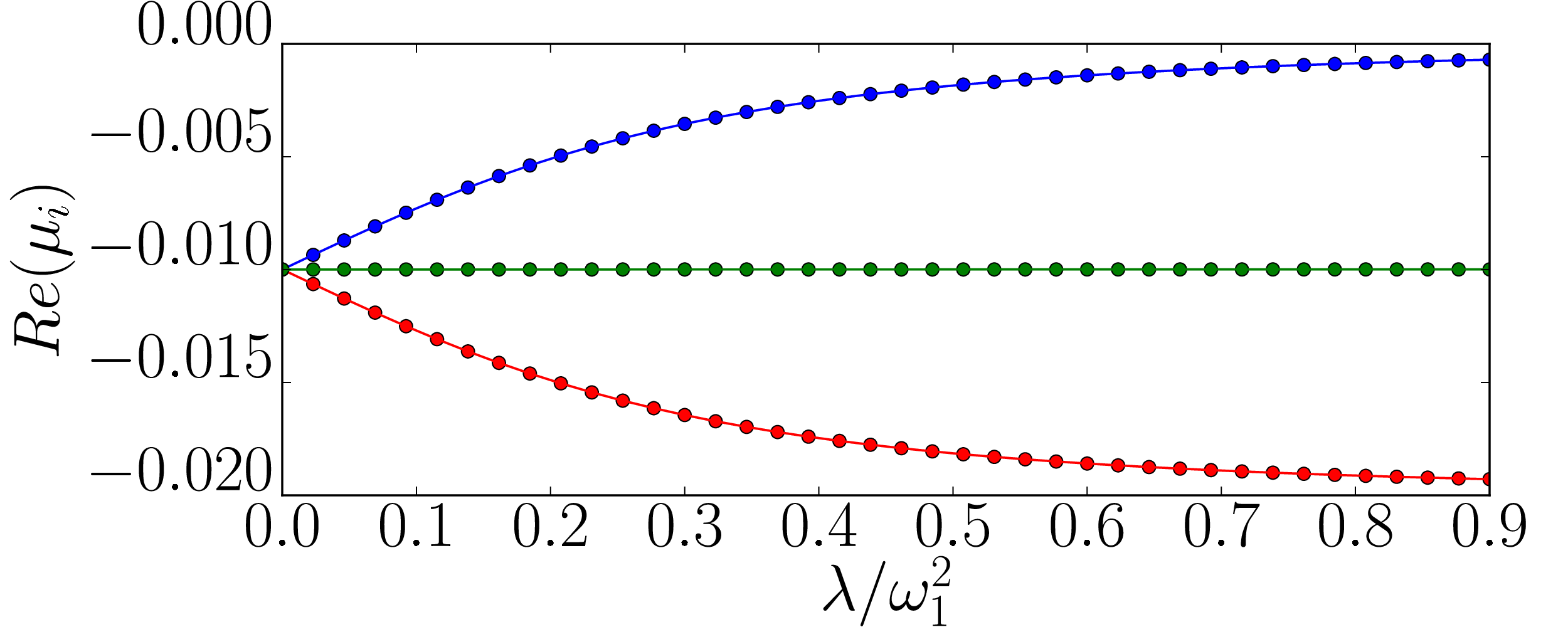}
\end{center}
\caption{ Rates $\tilde\Gamma_{++}^{CB}$, $\tilde\Gamma_{--}^{CB}$, from 
Eq.~(\ref{eq:decaysCB}) and
and $(\tilde\Gamma_{++}^{CB}+\tilde\Gamma_{--}^{CB})/2$ (dots) are compared with the (three different)  
real parts of the dynamical eigenvalues ${\rm Re}(\mu_{i}) $ (continuous line) in the case of a common bath for
$\omega_2/\omega_1 =1.31$.}
\label{fig3}
\end{figure}

The eigenmodes (\ref{eigenm1}) and (\ref{eigenm2}) diagonalize the Hamiltonian of the system $H_S$  but are still
indirectly coupled through the heat bath(s)  as seen from Eqs.  (\ref{eqofm})-(\ref{eqofm3}). This
means that the eigenmodes cannot be considered as independent channels for dissipation. Yet if we
rewrite their master equation in interaction picture, we can neglect fast oscillating terms, as usual in
the rotating-wave approximation, that is, eliminate exponents such as $e^{\pm i(\Omega_++\Omega_-)t}$
due to their highly oscillatory behavior in comparison with the overall slower dynamics
{ \cite{rivas,vasile}}. If we takethatwhich also rotate though more slowly, those such as
$e^{\pm i(\Omega_+-\Omega_-)t}$, and keep only nonrotating terms.  Finally, this procedure leads to an
effective total decoupling of the eigenmodes, which then dissipate independently to the heat bath(s)
with the decay rates  $\tilde\Gamma_{\pm\pm}^{CB}$ and $\tilde\Gamma_{\pm\pm}^{SB}$ [Eqs. (2) and (3)].
  In some sense this time averaging approximation can be seen as renormalizing all
dissipation coefficients having mixed indices $+-$ (and $-+$) to zero, hence rendering the master
equation as a tensor product of two independent evolutions. 
This could seem a bit too far fetched, but a comparison of the full
dynamics and this approximation seems to be quite accurate as clear from 
Fig.~\ref{fig3}, where we compare  $\tilde\Gamma_{\pm\pm}^{CB}$ and their average with the three
different values of ${\rm Re}(\mu_{i}) $. 

Inspection of these analytical expressions when varying system parameters confirms  that dynamical
eigenmode decays  for SB  all have similar real parts  (in general $\Gamma_{12}$ small  and
$\Gamma_{11}\simeq\Gamma_{22}$),  while  for a CB the decays can be significantly different  
(a factor $20$ in Fig.~\ref{fig3}). This difference between decay rates can be up to several orders
of magnitude for parameters where synchronization appears faster.
Synchronization is therefore linked to imbalanced
dissipation rates of the eigenmodes, allowing  the mode that survives longer to govern the dynamics. 
Within the discussed approximation its frequency is found to be $2\Omega_-$ with $\Omega_-$
 previously defined as the frequency of the eigenmode $X_-$. This is 
independent of bath coefficients and we find very good agreement with the exact frequency.

\begin{figure}[h]
\begin{center}
\includegraphics[width=9cm]{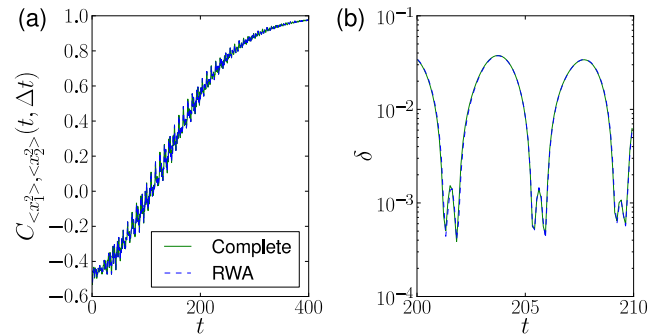}
\end{center}
\caption{  (a) Synchronization and (b)  discord obtained from the complete master equation 
(\ref{bas}) compared with the values obtained after the rotating-wave approximation as described in the text 
in the case of common bath for
$\omega_2/\omega_1 =1.4, ~\lambda=0.7\omega_1^2$.}
\label{fig_added}
\end{figure}

It can be easily seen that the rotating-wave approximation described in this appendix,
 neglecting all highly oscillatory
terms with exponents $e^{\pm i (\Omega_+ \pm\Omega_-)t}$, leads to (CB
and SB) master equations in the Kossakowski-Lindblad form \cite{breuer}. 
In particular, in the case of a common bath it can be found that
\begin{eqnarray}\label{basrw} 
\frac{d
\rho}{dt}&=&-i[H_S,{\rho}(t)]\\&-&\sum_{i=+,-} \frac{(\tilde{D}^{CB}_{ii}/\Omega_i)+\tilde{\Gamma}^{CB}_{ii} }{4}
[ A_i^\dag A_i \rho+\rho A_i^\dag  A_i-2  A_i \rho A_i^\dag]\nonumber\\&-&
\sum_{i=+,-}\frac{(\tilde{D}^{CB}_{ii}/\Omega_i)-\tilde{\Gamma}^{CB}_{ii}}{4}[
A_i  A_i^\dag \rho+\rho A_i A_i^\dag-2 A_i^\dag \rho A_i] \nonumber
\end{eqnarray} 
where $A_\pm=\sqrt{ \Omega_\pm/2}X_\pm+i/\sqrt{2  \Omega_\pm}P_\pm$.
In spite of formal differences  between Eqs. (\ref{bas}) and  (\ref{basrw})  we actually
find  very good agreement between their dynamical evolutions. In Fig. \ref{fig_added}
we show that, in the limit of weak 
coupling considered here, predictions for synchronization and discord are actually almost indistinguishable.
Maximum deviations in this case are at least two order of magnitude smaller than the represented quantities.
As
expected,  deviations increase for stronger system-environment coupling. 
The really weak dependence on the details of the master equation [Eqs. (\ref{bas}) and  (\ref{basrw})] when 
comparing the dynamical behavior of synchronization and quantum correlations between the oscillators
further strengthens the generality of our results.

\end{document}